\documentclass{article}
\usepackage[margin=1in]{geometry}
\usepackage[utf8]{inputenc}
\usepackage{natbib}
\usepackage{graphicx, caption}
\graphicspath{ {./graphics/} }
\usepackage{float}
\usepackage{algorithm}
\usepackage{algorithmic}
\usepackage{multicol}
\usepackage{amsmath}
\usepackage{caption}
\usepackage{subcaption}
\usepackage{tabto}
\usepackage{authblk}
\title{Optimizing Prediction of MGMT Promoter Methylation from MRI Scans using Adversarial Learning}
\author{Sauman Das}
\affil{Thomas Jefferson High School for Science and Technology}
\date{January 2022}

\begin{document}

\maketitle

\begin{abstract}
    Glioblastoma Multiforme (GBM) is a malignant brain cancer forming around 48\% of all brain and Central Nervous System (CNS) cancers. It is estimated that annually over 13,000 deaths occur in the US due to GBM, making it crucial to have early diagnosis systems that can lead to predictable and effective treatment. The most common treatment after GBM diagnosis is chemotherapy, which works by sending rapidly dividing cells to apoptosis. However, this form of treatment is not effective when the MGMT promoter sequence is unmethylated, and instead leads to severe side effects decreasing patient survivability. Therefore, it is important to be able to identify the MGMT promoter methylation status before deciding on the treatment plan. In this research, we aim to detect the methylation status through non-invasive magnetic resonance imaging (MRI) based machine learning (ML) models. This is accomplished using the Brain Tumor Segmentation (BraTS) 2021 dataset, which was recently used for an international Kaggle competition. We developed four primary models - two radiomic models and two CNN models - each solving the binary classification task with progressive improvements. We built a novel ML model termed as the \textit{Intermediate State Generator} which was used to normalize the slice thicknesses of all MRI scans. With further improvements, our best model was able to achieve performance significantly ($p<0.05$) better than the best performing Kaggle model with a 6\% increase in average cross-validation accuracy. This improvement could potentially lead to a more informed choice of chemotherapy as a treatment option, prolonging lives of thousands of patients with GBM each year. 
\end{abstract}

\section{Introduction}

\tab A recent survey conducted by GLOBOCAN in 185 countries, estimates that there were 308,102 new cases of brain cancer and 251,329 deaths in 2020 alone in 185 countries~\citep{sung2021global}. \emph{Glioblastoma Multiforme (GBM)} is one of the most stubborn and aggressive forms of malignant brain cancer, with very few effective treatment options. The World Health Organization has classified it as a grade IV, the highest grade brain tumor, accounting for 48\% of all primary malignant brain tumors~\citep{ostrom2014cbtrus}. More than 13,000 Americans are expected to receive a GBM diagnosis each year. It is also estimated that more than 10,000 Americans will die from GBM every year.

The  typical recommended treatment for GBM patients include surgical removal of the tumor followed by radiotherapy and chemotherapy. With this standard care, the patients have an median survival time of 15 months as compared to a mere 4 months if left untreated once diagnosed~\citep{bleeker2012recent}. Chemotherapy, a standard and often used treatment option, works by killing rapidly dividing cells but cannot always differentiate between tumor cells from normal cells. This can result in adverse side effects~\citep{macdonald2009chemotherapy}~\citep{taal2015chemotherapy}. According to the American Cancer Society, for GBM treatment, these side effects may include peripheral neuropathy, which includes symptoms causing damage to the central nervous systems and impaired body movements, significantly deteriorating patient quality of life. Therefore, before the patient starts chemotherapy treatment, it is critical to know if it would be effective in curbing the disease progress.

Chemotherapy relies on Temozolomide (TMZ) which places a side product on a Guanine molecule of the DNA. This additional molecule prevents the synthesis of new DNA and leads to apoptosis of the cancer cells. However, an enzyme, known as $\text{O}^6$-methylguanine DNA methyltransferase (MGMT), can remove the additional molecule. If the MGMT protein is transcribed, then cells will not be led to apoptosis because of the absence of the additional molecule and chemotherapy will be ineffective. Sometimes, however, the promoter region of the MGMT gene is methylated and that prevents transcription of the enzyme. As a result, chemotherapy will likely be effective, since the side product molecule will remain attached to the Guanine molecule. 

Unfortunately Temozolomide has severe side effects, especially in patients with pre-existing comorbidities. Hence, detection of MGMT promoter methylation status may help tailor the best Temozolomide dosage or schedule for the GBM patient. The MGMT promoter methylation status can be found through invasive surgeries; however, this is time-taking and can lead to side effects caused by the surgery itself. However, with the increasing amount of biomedical imaging data along with recent advances in deep learning, the process of MGMT promoter methylation detection can be largely simplified. This work aims to discover if MGMT promoter methylation can be detected from MRI scans and reveal certain biomedical markers extracted from images that strongly correlate with the methylation status. This approach could potentially result in a non-invasive and more efficient form of prognosis leading to better treatment for GBM.

In this work, we aim to improve the accuracy of MGMT promoter methylation prediction, but achieve this using data that is noisy and more representative of what would be provided in a real medical setting. We begin by summarizing prior research done on this topic in Section~\ref{relatedworks}. Then, we will go into detail about methodologies used in this work including the explanation of an original preprocessing strategy in Section~\ref{enhance}. Furthermore, we train and test 4 different models and summarize their results. Finally, we discuss the results shown in this paper with respect to work from prior research.

\section{Existing Research}\label{relatedworks}

During the summer of 2021, the Radiological Society of North America hosted a prestigious Kaggle competition to encourage renewed research on MGMT promoter methylation detection using data science and machine learning techniques. The challenge received over 1,500 submissions from researchers around the world. Firas Baba used an ML model known as \emph{ResNet} and was declared as the eventual winner of the competition. The model generated the highest area under the ROC curve score of 0.62714. 

\begin{figure}[H]
    \centering
    \captionsetup{width=4in}
    \includegraphics[width=6in]{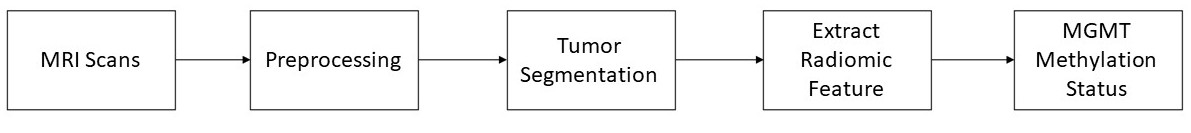}
    \caption{Overview of Radiomic Feature Extraction Approach}
    \label{img1}
\end{figure}

Prior to this Kaggle challenge, researchers have primarily followed two distinct approaches to detect MGMT Promoter methylation. The first approach is outlined in Figure~\ref{img1}.  


A brain tumor segmentation model is created to identify the region of interest (ROI). Then, radiomic features, such as tumor location, tumor shape, and several histogram and gray level coocurrence matrix (texture descriptors) are extracted. These features are then processed by machine learning models to predict the methylation status.

\begin{figure}[H]
    \centering
    \captionsetup{width=4in}
    \includegraphics[width=4.5in]{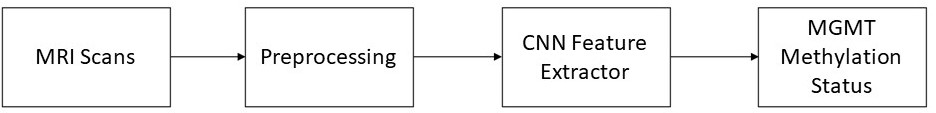}
    \caption{Overview of CNN Approach}
    \label{cnn}
\end{figure}

During the Kaggle competition, some researchers found that radiomic feature extraction did not provide the most optimal results. Instead, they followed an alternate approach which involved using Convolutional Neural Networks (CNNs) rather than extracting radiomic features. This method does not require tumor segmentation masks which the previous method relied upon. The CNN method is outlined in Figure~\ref{cnn}.  

\cite{ahn2014prediction} was one of the earliest to study and analyze the MGMT promoter methylation from MRI scans. They studied the MRI scans of 43 patients diagnosed with GBM and were provided with the methylation status for each patient. They extracted features including the rate transfer coefficient, apparent diffusion coefficient, and fractional anisotropy from a defined ROI. Using Chi-Square and Mann-Whitney tests, they found the correlation between the extracted imaging features and the MGMT promoter methylation status. They achieved sensitivity and specificity values of 56.3\% and 85.2\%, respectively.

\cite{sasaki2019radiomics} did a more thorough analysis in describing the effectiveness of radiomic features for methylation status detection. They extracted 489 texture features from the ROI along with location features. Through Principal Component Analysis (PCA), they were able to identify 22 features that contributed most to the MGMT promoter methylation. They were able to achieve a high accuracy of 67\% with their LASSO regression model.

\cite{han2018mri} followed the Convolutional Neural Network (CNN) feature extraction method (Figure~\ref{cnn}). They proposed a novel bi-directional CNN architecture to predict MGMT promoter methylation. The benefit of using this type of network is that any number of MRI slices can be passed through due to the recurrent nature of the network. In contrast, a 3-dimensional CNN would require a uniform number of slices to work. They tested their model on 21 patients and achieved a 62\% accuracy.


\section{Materials and Methods}

\subsection{Dataset}

\tab For this research We utilized the MRI datasets provided by the Brain Tumor Segmentation (BraTS) Challenge in 2020 and 2021. To build a segmentation model to detect the tumour region, BraTS 2020 Challenge dataset was used. 369 MRI scans were provided in four modalities, namely Fluid-attenuated inversion recovery (FLAIR), T1-weighted (T1w), T2-weighted (T2w), and T1-weighted contrast enhanced (T1wCE). Both the scans and the segmentation masks were given in the NIfTI file format and in the coronal orientation. The  masks provided four class labels including non-tumor, non-enhancing tumor core, peritumoral edema, and enhancing tumor. For our research, we combined the latter three classes since we were only concerned with the general tumor region.

In 2021, the Radiological Society of North America (RSNA) along with the Medical Image Computing and Computer Assisted Intervention (MICCAI) Society introduced the MGMT promoter methylation detection aspect to the BraTS challenge hosted on Kaggle. They provided a pre-selected dataset of 585 MRI scans in the same four modalities as in 2020. The scans were collected from various institutions with different equipment in order to represent the diverse clinical practices across the world~\citep{baid2021rsna}. For the classification task, we used these dataset which were in the DICOM format labelled with their methylation status. The MGMT methylation status was confirmed by laboratory assessment of the surgical brain tumor specimen. The four modalities remained the same from the previous dataset described, but the T1w scans were not used due to consistently large thicknesses contributing to more noise than information.

Unlike many prior research on methylation detection, our MRI scans were very \textit{messy} containing varying slice thicknesses and orientations. This messiness is typically observed in real-world clinical data. Although preprocessing of messy data is challenging, an autonomous deep learning model should apply adequate correction to avoid any intervention from the medical practitioners. The practitioner should be able to specify these messy MRI scans to the ML model in order to generate the methylation status.

\subsection{Overview of Models}
\begin{figure}[H]
    \centering
    \includegraphics[width=6.6in]{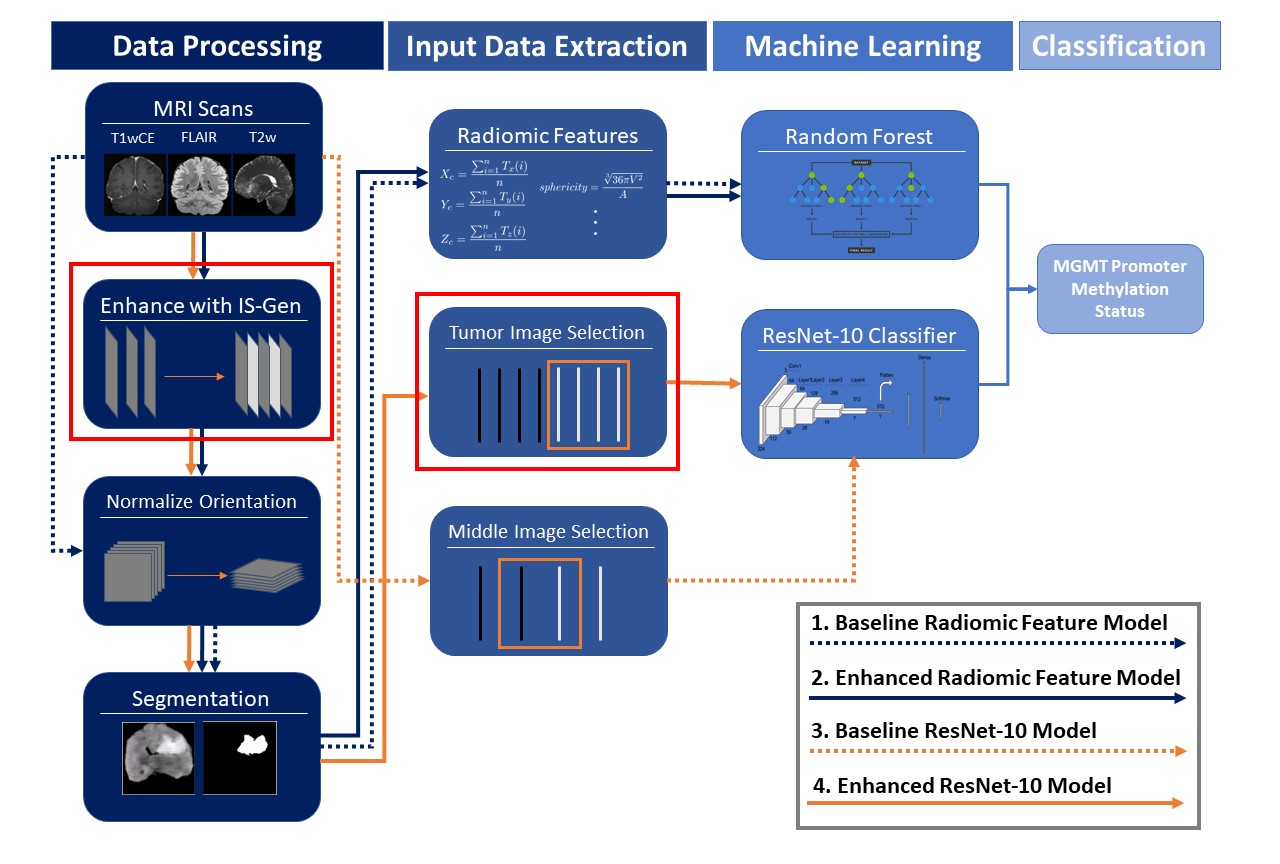}
    \caption{Overview of ML modeling approaches for MGMT detection.}
    \label{fig:ModelOverview}
\end{figure}

\tab The objective of this research is to improve the accuracy of MGMT prediction using ML models for messy real-world MRI data. For this purpose we investigated potential improvements that can be introduced at various stages of the processing to arrive at the final methylation decision. Figure~\ref{fig:ModelOverview} provides an overview of the four specific models we developed to study and compare the performance using the same messy dataset. The first two models relied on radiomic feature based approach as adopted by many of researchers recently. Unlike these models, the latter two models used CNN directly without explicit feature extraction. The first and third models are mere replications of known prior approaches. The third model in particular has gained a lot of recognition as the winner of the 2021 Kaggle challenge.

The second and fourth models are based on the basic prior approaches of first and third models respectively. These new models exhibited improved performance when the messy data is first corrected and then used for classification. The correction is applied by a new proposed method called \textit{Intermediate State Generator} or \textit{IS-Gen} which fills in missing images in a particular MRI scan with images imputed from adjacent real images. This way all MRI scans are normalized to have the same number of images with equal thickness. The normalization is fully complete when all the images are rotated to be in the coronal plane. Rotating the images after application of IS-Gen prevents undue blurring of the rotated images. 

The traditional CNN approach of third model considers the 64 slices around the middle image for use by CNN for classification. One potential issue with this approach is the assumption that the tumor region is more or less centrally located. Furthermore, these 64 slice selection does not consider the absence of slices in heterogeneous messy MRI scans. The fourth and final model corrects these issues by first applying IS-Gen to fill in the missing images, then finding the tumor region to detect the location of maximal presence of the disease. The final 64 slices are selected surrounding the maximal presence of the tumor for the CNN classification

\subsection{Baseline Radiomic Feature Based Model}\label{rad_baseline}

\tab In Section~\ref{relatedworks}, two approaches to achieve MGMT promoter methylation prediction were discussed. For the first model, we use the radiomic feature extraction approach. The first step was to build a segmentation model to define the tumor ROI. We used a 2D U-Net that was originally designed by Rastislav\footnote{https://www.kaggle.com/rastislav/3d-mri-brain-tumor-segmentation-u-net}. We optimized the model for better use in the complete pipeline. The primary modification we made was to combine the three non-background classes into one tumor class. Furthermore, we cropped all MRI scans to the smallest rectangular prism containing the full brain (i.e. removed all black padding). 

The U-Net was trained for 35 epochs on MRI scans provided in the coronal plane. We monitored the Intersection Over Union (IOU) and dice coefficient which are frequently used metrics to measure image segmentation performance. Given a predicted mask $A$ and labelled region $B$, the following equations can be used to calculate IOU and the dice score.

\[
\text{IOU} = \frac{|A \cap B|}{|A \cup B|}
\]
\[
\text{Dice Coefficient} = \frac{2|A \cap B|}{|A| + |B|}
\]

The final U-net exhibited validation IOU and dice scores of 0.832 and 0.626, respectively. 


The segmentation model was then applied to the MRI scans in the BraTS 2021 dataset which contained methylation status labels. The model was applied to each individual slice reoriented into the Coronal plane in the MRI scan because of the 2D U-net architecture. Once masks were defined, we began extracting radiomic features. Radiomics refers to the idea of extracting features from biomedical images such as computed tomography, positron emission tomography, or MRIs in order to build ``models relating image features to phenotypes or gene–protein signatures"~\citep{kumar2012radiomics}. We extracted 39 features from 3 modalities of MRI scans using the Python library known as PyRadiomics~\citep{van2017computational}. We included both first-order features such as energy and entropy of the tumor region and included some texture features using the Gray Level Co-occurrence Matrix (GLCM) and Run Length Matrix (GLRLM). Additionally, we calculated the tumor region centroid coordinates along with shape features of the tumor. The centroid coordinates were treated as 3 separate features which were calculated using the formulas below where $T$ represents the $x$, $y$, or $z$ coordinate of the $i$th pixel located in the tumor and $n$ is the total number of pixels in the tumor.

\begin{align*} 
X_c = \frac{\sum_{i=1}^{n} T_x(i)}{n} && Y_c = \frac{\sum_{i=1}^{n} T_y(i)}{n} && Z_c = \frac{\sum_{i=1}^{n} T_z(i)}{n}
\end{align*}

We used another 3 shape features including approximate mesh volume (V), surface area (A), and sphericity (closeness to a sphere) using PyRadiomics. Out of the 39 total features extracted, the 7 features which produced the highest accuracies during cross-validation were the following: $X_c$, $Y_c$, $Z_c$, Volume, Surface Area, Sphericity, and Total Energy calculated from the FLAIR modality. 

After doing preliminary experimentation with multiple machine learning models including Decision Trees, Random Forests, AdaBoost, Support Vector Machines, and Extreme Gradient Boosting (XGBoost), we further fine-tuned the Random Forest since it showed the most promising results. All these models were implemented using the Scikit Learn package. The following set of hyperparameters were experimented with for the Random Forest using a Grid Search:

\begin{table}[h!]
\centering
\begin{tabular}{||l | l||} 
 \hline
 \textbf{Hyperparameter} & \textbf{Values} \\
 \hline\hline
 n-estimators & [\textbf{50}, 150, 200, 250, 300]\\ 
 \hline
 criterion & [\textbf{`gini'}, `entropy'] \\
 \hline
 max-depth & [\textbf{10}, 15, 20, 25, 30, 35, 40, 45, 50] \\
 \hline
 max-leaf-nodes & [5, 10, 15, 20, 30, \textbf{None}] \\
 \hline
 class-weight & [`balanced', \textbf{None}] \\ 
 \hline
 min-samples-split & [2, 4, \textbf{6}, 8]\\
 \hline
\end{tabular}
\caption{Parameter grid for random forest fine-tuning.}
\end{table}

The bolded values produced the optimal results. A 5-fold Cross Validation (CV) was conducted for concrete model evaluation. The average results are shown below.

\begin{table}[h!]
\centering
\begin{tabular}{||c | c | c | c | c | c||} 
 \hline
 \textbf{Accuracy} & \textbf{Sensitivity} & \textbf{Specificity} & \textbf{PPV} & \textbf{NPV} & \textbf{AUROC} \\
 \hline\hline
 0.564102564 & 0.678241587 & 0.418761311 & 0.615846663 & 0.678241587 & 0.548501449\\ 
 \hline
\end{tabular}
\caption{Average Cross Validation Results for Baseline Radiomic Feature Extraction Model.}
\end{table}

One possible explanation for the poor results was the varying slice thickness and general noise prevalent in the MRI scans. The radiomic approach heavily relies on the accuracy of segmentation and clarity of images because radiomic features are extracted based on pixel values. \cite{thrower2021effect} show that a MRI slice thickness greater than 1 mm effects segmentation of brain lesions. Therefore, being able to have MRI scans with minimal slice thickness could potentially provide more tumor information resulting in a better radiomic feature extractor.  

However, we cannot rely on alternative data sources such as TCIA which provide higher quality MRI scans. Data provided in real-world medical conditions will not necessarily be processed to the way that benefits our model. Most likely, we will be faced with varying slice thicknesses and orientations just as the BraTS 2021 dataset presents. This problem led to the creation of a new architecture used for preprocessing called the Intermediate State Generator.

\subsection{Intermediate State Generator (IS-Gen)}\label{enhance}
In this paper, we consider the orientation and slice thickness problems inherent in the messy data. Orientation refers to the plane in which an MRI scan is provided. There are three distinct orientations through which MRI scans can be interpreted, namely sagittal, axial, and coronal. The diagram below shows which plane each of the three orientations refer to. 

\begin{figure}[H]
    \centering
    \includegraphics[width=2in]{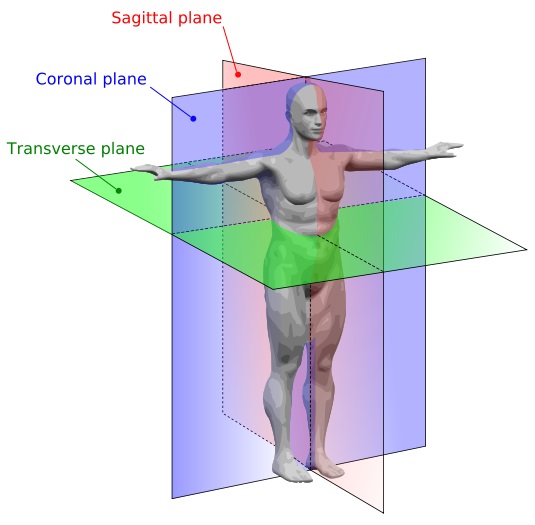}
    \caption{Three different MRI planes. The axial plane is sometimes referred to as the transverse plane. Source: My-MS.org}
    \label{fig:orientations}
\end{figure}

Although orientation does not have to remain constant when passed through a model, the model can benefit by removing the orientation variable to learn from. Furthermore, the segmentation dataset provides all images in the coronal orientation which results in more accurate tumour region segmentation for the classification images in that orientation.

The process of modifying orientations to be constant throughout the data directly relates to problems caused by slice thickness. A larger slice thickness indicates the availability of less 2D slices to represent the brain volume. We can see the consequence of changing orientation with large slice thicknesses in the following figure.  

\begin{figure}[H]
     \centering
     \begin{subfigure}[b]{1.5in}
         \centering
         \includegraphics[width=\textwidth]{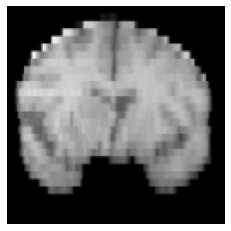}
         \caption{}
         \label{fig:blurry1}
     \end{subfigure}
     \begin{subfigure}[b]{1.5in}
         \centering
         \includegraphics[width=\textwidth]{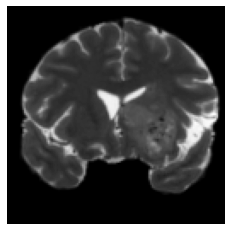}
         \caption{}
         \label{fig:crisp1}
     \end{subfigure}
     \caption{Image (a) is created by changing orientation of an axial T1w MRI with 33 total slices. Image (b) is created by changing orientation of a sagittal T2w MRI with 274 slices. The images are presented in the figure in the coronal plane.}
\end{figure}

Figure~\ref{fig:blurry1} is very blurry and pixelated because only 33 slices were originally provided in the axial orientation. The minimal amount of slices is not enough to produce a clear image of the brain when reinterpreted into the coronal orientation. The image would not be good enough for a segmentation model to locate the tumor due to extremely low level of detail. Because there were several T1w images with large slice thickness, we excluded them from our dataset. However, Figure~\ref{fig:crisp1} also shows an image that has been reinterpreted into the coronal plane and the level of detail that is preserved is greatly improved because of a significantly lower slice thickness.

In order to get automated MGMT promoter methylation detection to the level of deployment, the input should be as generalized as possible. The goal is to be able to provide a prognosis given any form of the MRI scans in the three modalities (i.e. T1wCE, T2w, FLAIR). Consequently, we developed a preprocessing algorithm that could adjust the slice thickness by inserting \textit{synthetic} MRI slices based on the real MRI sequence. We hypothesized that this would be able to provide more accurate segmentation and also increase accuracy of texture analysis. Three millimeter slice thickness is generally what is required for accurate texture analysis~\citep{savio2010effect}.

We assume that this task is reasonable due to biological continuity that naturally exists. In other words, there must be well-defined relationships between 3 consecutive slices of an MRI sequence provided their distance between each other is not too large. Our goal was to create synthetic slices that provide \textbf{more information} than what existed previously. So, once tested, it should be more accurate than what we call the ``copy imputer": the process of filling in intermediary slices by duplicating adjacent slices that already exist.

\subsubsection{Initial Model Architecture}\label{construct}

\tab We planned to create synthetic slices with a model called the Intermediate State Generator (IS-Gen). The IS-Gen takes in 2 images $x_1$ and $x_2$ as input, and attempts to produce the MRI slice belonging between the two inputs denoted by $\hat{y}$. Our initial model contained two separate convolutional feature extractors that each took in a 256x256 sized MRI scan and extracted 2 feature maps of size 16x16x64. These feature maps were then concatenated and passed through upsampling convolutions resulting in a 256x256 sized reconstruction of the original image. 

\begin{figure}[H]
    \centering
    \includegraphics[width=6in]{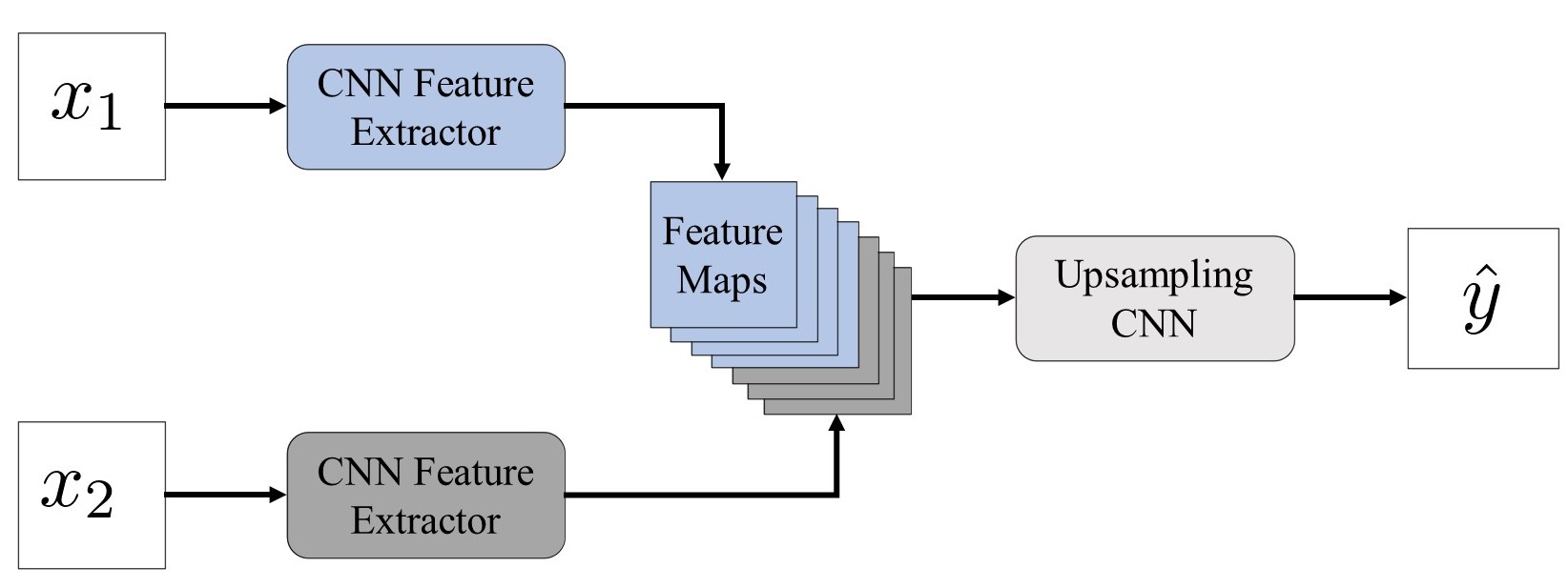}
    \caption{Initial IS-Gen model outline.}
    \label{fig:isgen-v1}
\end{figure}

The model was trained using a supervised algorithm by taking sequences of 3 images separated by an equal width and using the second image ($y$) as the label. The spacing between the images passed through the model was randomly selected during training to provide a variety of slice thicknesses since the model would need to work on a variety of slice thicknesses when used during preprocessing. 

We defined the loss function as the mean squared error in order to achieve the most accurate reconstruction. 

\[
\mathcal{L}_{RL}(y, \hat{y}) = (y-\hat{y})^2
\]

However, due to overrepresentation of black pixels, the model began returning pure black images providing zero additional information. To combat this issue, we introduced a piecewise loss function that adapted to the pixel values. 

\[
    \mathcal{L}_{RL}(y, \hat{y}) = 
    \begin{cases} 
      \hat{y}^2 & y = 0 \\
      5*(y-\hat{y})^2 & y \neq 0
   \end{cases} 
\]

Training this model allowed the model to learn fine details of the brain. After training the model for 40 epochs, the following results were achieved. 
\begin{figure}[H]
     \centering
     \begin{subfigure}[b]{1.8in}
         \centering
         \includegraphics[width=\textwidth]{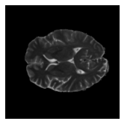}
         \caption{}
         \label{fig:t1}
     \end{subfigure}
     \begin{subfigure}[b]{1.8in}
         \centering
         \includegraphics[width=\textwidth]{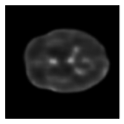}
         \caption{}
         \label{fig:blurry}
     \end{subfigure}
     \caption{Example results from mean squared error training. Image (a) represents the actual intermediary label. Image (b) is the output of the IS-Gen with two adjacent slices provided as input. Notice the blurriness that is present in the image.}
\end{figure}

The images are able to maintain relative shape and color shade of the scan; however, the detail in the center is completely blurred. A segmentation model would likely not be able to accurately extract the tumor region from the generated slice because there is very little resemblance to a brain at this stage. 

\subsubsection{Discriminator Introduction}

In order to combat the problems with the current synthetic imputations, we added a new adversarial component to the IS-Gen called the discriminator. A discriminator, in adversarial models, is typically used to detect whether the produced image/result is real or fake. During training, the generator (i.e. component described in Figure~\ref{fig:isgen-v1}) attempts to create a synthetic image that is close to the label defined by $\mathcal{L}_{RL}$, but it also tries to create an image that can trick the discriminator into thinking it is real. Simultaneously, the discriminator is learning how to discriminate between real and fake images. The discriminator is a basic binary classification CNN with 3 convolutional layers, a fully connected layer, and a single neuron sigmoid output layer.

\begin{figure}[H]
    \centering
    \includegraphics[width=6in]{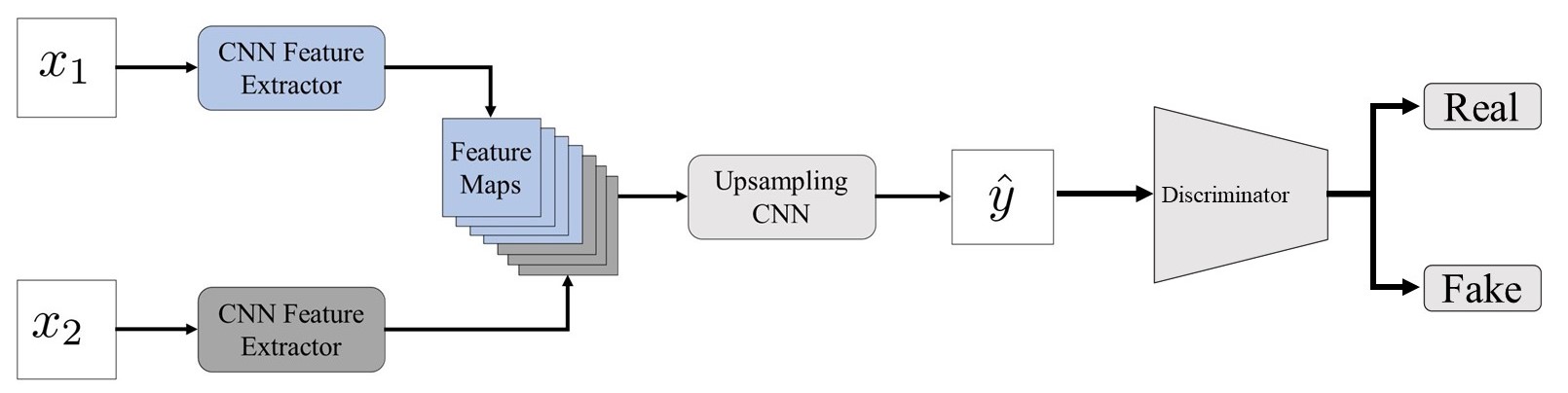}
    \caption{Full IS-Gen model with discriminator to ensure crisp image creation.}
    \label{fig:isgen-v1}
\end{figure}

The model architecture has some similarities to a Generative Adversarial Network (GAN)~\citep{goodfellow2014generative}, but adds a reconstruction loss to the generated image. Furthermore, this is trained in a supervised manner whereas GANs use an unsupervised training algorithm. 

\subsubsection{IS-Gen Training}

Our initial approach towards training the IS-Gen was similar to that of the GAN training algorithm. The sequences of three images ($x_1, x_2, y_g$) are selected in the same way described in Section~\ref{construct}. Then, the generator $G$ produces an intermediary slice prediction $\hat{y_g}$. The discriminator takes $\hat{y_g}$ and $x_1$ as input and produces a binary prediction for each of the images. The goal for the discriminator is to predict each generated image as belonging to class 0 and real image belonging to class 1. It is trained with the following loss function:

\[
\mathcal{L}_D(y_d, \hat{y_d})=-\frac{1}{N}\sum_{i=1}^{N} y_{di}\cdot \log(\hat{y_{di}}) + (1-y_{di})\cdot \log(1-\hat{y_{di}})
\]

In our experimentation, the value of $N$ was 2 because we passed in exactly one synthetic and one real image in each batch. Additionally, the generator loss function was expanded to include the adversarial component. The labels for the real vs. synthetic images were inverted for the generator because the generator's task is the opposite of the discriminators. The following equation represents the loss function for the generator:

\[
\mathcal{L}_G(y_g, \hat{y_g}, \overline{y_d}, \hat{y_d}) = \mathcal{L}_{RL}(y, \hat{y}) + \lambda \mathcal{L}_{D}(\overline{y_d}, \hat{y_d})
\]

Note that we added the extra $\lambda$ factor to the adversarial loss function term. Originally we set $\lambda=1$, but the model was not able to learn useful information. The generator was not able to learn fast enough relative to the discriminator which again resulted in the creation of black images. Consequently, we introduced the regularization term and set $\lambda=0.03$ to scale down the impact of the discriminator. The adversarial training algorithm is shown below: 

\begin{algorithm}[H]
\label{alg:algo1}
\caption{Adversarial IS-Gen Training Step}
\begin{algorithmic}[1]
\FORALL{$x_1, x_2, y_g\in Data$}
\STATE  $\hat{y_g} \leftarrow G(x_1, x_2)$
\STATE  $\hat{y_d} \leftarrow D(\hat{y_g}, x_1)$ 
\STATE  $y_d \leftarrow [0, 1]$ 
\STATE  Update: $[w_g \leftarrow w_g-\eta \nabla \mathcal{L}_{G}(y_g, \hat{y_g}, \overline{y_d}, \hat{y_d})]$
\STATE  Update: $[w_d \leftarrow w_d-\eta \nabla \mathcal{L}_{D}(y_d, \hat{y_d})]$
\ENDFOR
\end{algorithmic}
\end{algorithm}

These modifications successfully decreased the blurriness of the IS-Gen output. However, after 10 epochs of purely adversarial training (i.e. training with $\lambda=0.03$), the model began to decrease in accuracy as measured by the reconstruction loss.

\begin{figure}[H]
     \centering
     \begin{subfigure}[b]{1.8in}
         \centering
         \includegraphics[width=\textwidth]{graphics/true.png}
         \caption{}
         \label{fig:t1}
     \end{subfigure}
     \begin{subfigure}[b]{1.8in}
         \centering
         \includegraphics[width=\textwidth]{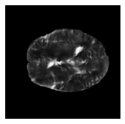}
         \caption{}
         \label{fig:only_ad}
     \end{subfigure}
     \caption{Sample synthetic image generated after 10 epochs of adversarial training. We can clearly see decreased accuracy compared to Figure~\ref{fig:blurry}, but the blurriness has significantly reduced.}
\end{figure}

We used one final modification to improve the training of the IS-Gen called ``On-Off Training". During On-Off training, the model does not consistently learn using adversarial training; instead, it switches back and forth between adversarial and non-adversarial training to increase clarity while maintaining pixelwise accuracy. 

For the first 5 epochs, the model was trained with $\lambda=0$ and gradients were not backpropagated through discriminator enabling the generator to get a head start (Algorithm 2). For the following 5 epochs, we again set $\lambda=0.03$ and allowed backpropagation of gradients through the discriminator (Algorithm 1). This 10 epoch pattern was repeated 10 times resulting in 100 total epochs of training.

\begin{algorithm}[H]
\label{alg:algo2}
\caption{Non-Adversarial IS-Gen Training Step}
\begin{algorithmic}[1]
\FORALL{$x_1, x_2, y_g\in Data$}
\STATE  $\hat{y_g} \leftarrow G(x_1, x_2)$
\STATE  Update: $[w_g \leftarrow w_g-\eta \nabla \mathcal{L}_{RL}(y_g, \hat{y_g})]$
\ENDFOR
\end{algorithmic}
\end{algorithm}

Using this process prevented the model from losing too much clarity and maintained accuracy in its predictions.


\begin{figure}[H]
     \centering
     \begin{subfigure}[b]{1.8in}
         \centering
         \includegraphics[width=\textwidth]{graphics/true.png}
         \caption{}
         \label{fig:t1}
     \end{subfigure}
     \begin{subfigure}[b]{1.8in}
         \centering
         \includegraphics[width=\textwidth]{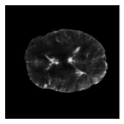}
         \caption{}
         \label{fig:t2}
     \end{subfigure}
     \caption{The produced image (b) is significantly more accurate than Figure~\ref{fig:only_ad} and maintains clarity unlike Figure~\ref{fig:blurry}. Image (a) is the provided intermediary image.}
\end{figure}

This process was applied to three different IS-Gen models that were specific to 3 modalities of MRI scans, namely FLAIR, T1wCE, and T2w.

The updated results show that both shape and clarity have been conserved to an extent. We can also see that the color distribution remains relatively accurate indicating that a segmentation model could also work for the generated slices. 

We wanted to confirm that the model was contributing more information than what already existed in the MRI sequence. Consequently, we compared the IS-Gen to a basic ``copy imputer". The copy imputer filled in missing slices by simply adding a copy of an adjacent slice that already existed, therefore contributing no new information, yet filling in slices with some level of accuracy. 


\begin{table}[H]
\centering
\begin{tabular}{|| l | c | c | c ||} 
 \hline
 & \textbf{T1wCE} & \textbf{FLAIR} & \textbf{T2w} \\
 \hline\hline
 \textbf{IS-Gen} & 4.820610455 & 6.069807723 & 6.204056403
 \\ 
 \hline
\textbf{Copy Imputer} & 22.25231586 & 27.96860923 & 28.92950342 \\
\hline
\end{tabular}
\caption{Average Mean Absolute Pixelwise Error from 1,250 test cases. All pixel values were scaled between 0-255 before score calculation.}
\end{table}

The results of the IS-Gen were significantly better ($\alpha=0.05$) than naively copying adjacent slices. This shows that the IS-Gen was able to contribute information that did not previously exist while maintaining a relatively low average pixelwise error.  

\subsubsection{Applying the IS-Gen}

\tab For all of the models that we design, we resize the MRI scans to contain 128 slices. Based on provided metadata, 128 slices in the coronal plane represents a slice thickness of approximately 2.5 mm. 

The goal of the IS-Gen is to standardize the slice thickness of the MRI scans by providing synthetic images to scans with a large slice thickness. For scans that contain less than 128 slices in the provided orientation, the IS-Gen was used iteratively to impute intermediary slices. During the first round of imputation, existing adjacent pairs of slices were used as input to the IS-Gen to impute new slices (Figure 11). 

\begin{figure}[H]
     \centering
     \begin{subfigure}[b]{1in}
         \centering
         \includegraphics[width=\textwidth]{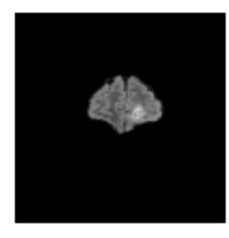}
         \caption{}
         \label{fig:ex1}
     \end{subfigure}
     \begin{subfigure}[b]{1in}
         \centering
         \includegraphics[width=\textwidth]{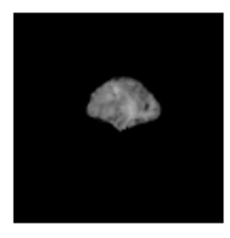}
         \caption{}
         \label{fig:ex2}
     \end{subfigure}
    \begin{subfigure}[b]{0.99in}
         \centering
         \includegraphics[width=\textwidth]{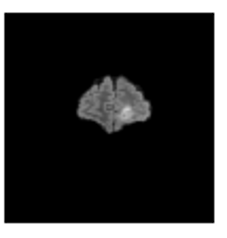}
         \caption{}
         \label{fig:ex2}
     \end{subfigure}
     \label{fig:impute}
     \caption{Image (a) and (c) are two adjacent slices in ad MRI scan. Image (b) represents the middle slice produced by passing in (a) and (c) as input to the IS-Gen trained with On-Off training.}
\end{figure}

If the first round of imputation did not result in at least 128 slices, the same process was repeated with the current MRI scan until it has all 128 slices in the provided plane. If the imputation resulted in more than 128 slices, exactly 128 slices were selected in uniform from the existing scan. Once the slices were selected, the MRI scan was reoriented into the coronal plane, if necessary. The final MRI scan dimensions were 128x128x128.

\subsection{Enhanced Radiomic Feature Based Model}\label{enh_radiomic}

We hypothesized that the IS-Gen could contribute new information to the MRI allowing for a more detailed and representative set of radiomic features. Especially for the shape features, we believe that the IS-Gen could be used to preprocess and enhance the provided MRI scans to improve the quality of these features. In this paper, we will refer to an MRI scan that has been preprocessed by the IS-Gen as \textit{enhanced}. 

The pipeline for radiomic feature extraction presented in Section~\ref{rad_baseline} was reused to build another model with one modification. We enhanced all the MRI scans before passing them through the segmentation model and all other factors remained constant. The radiomic features were now going to be extracted from MRIs with equal slice thickness and additional information. 

Similar to the previous baseline radiomic feature model, 39 features were extracted but the 6 shape and location features provided the optimal preliminary results when trained with the random forest classifier. We performed a grid search with the hyperparameters presented in Table 1 and the final hyperparameters are presented in the table below. 

\begin{table}[h!]
\centering
\begin{tabular}{||l | l||} 
 \hline
 \textbf{Hyperparameter} & \textbf{Optimal Value} \\
 \hline\hline
 n-estimators & 200\\ 
 \hline
 criterion & `gini' \\
 \hline
 max-depth & 10 \\
 \hline
 max-leaf-nodes & 5 \\
 \hline
 class-weight & None \\ 
 \hline
 min-samples-split & 4\\
 \hline
\end{tabular}
\caption{Parameter grid for random forest fine-tuning.}
\end{table}

The results of the 5-fold CV for the enhanced radiomic feature extraction model are shown in the table below. 

\begin{table}[h!]
\centering
\begin{tabular}{||c | c | c | c | c | c||} 
 \hline
 \textbf{Accuracy} & \textbf{Sensitivity} & \textbf{Specificity} & \textbf{PPV} & \textbf{NPV} & \textbf{AUROC} \\
 \hline\hline
 0.606837607 & 0.770721608 & 0.401783756 & 0.636609161	& 0.566618256 & 0.586252682
\\ 
 \hline
\end{tabular}
\caption{Average Cross Validation Results for Enhanced Radiomic Feature Extraction Model.}
\end{table}

These results show a borderline statistically significant improvement ($p=0.049$) from the baseline radiomic model as shown by a 2-way ANOVA with 5 replications (5 folds of CV); however, the results need significant improvement because certain metrics such as the specificity are around 40\% for both models. The increase in accuracy can primarily be credited to the 10\% increase in sensitivity. 

Although the radiomic feature extraction process was not the most successful in terms of performance, it showed that the IS-Gen preprocessing has the potential to contribute additional information towards the methylation status prediction because that was the only variable modified from the baseline model.

\subsection{Baseline ResNet-10}

As discussed in Section~\ref{relatedworks}, two modeling approaches are commonly followed to detect MGMT promoter methylation prediction, viz. radiomic feature extraction and CNN feature extraction. As described in the previous section, radiomic feature extraction based approach resulted in only modest  improvement in MGMT detection accuracy. We therefore proceeded to experiment with the alternative latter approach. 

As mentioned earlier, we utilized the dataset used in the Kaggle competition hosted during the summer of 2021. Incidentally, the winner, Firas Baba, described his results in a Kaggle discussion post\footnote{https://www.kaggle.com/c/rsna-miccai-brain-tumor-radiogenomic-classification/discussion/281347}. At the time of our research, there did not exist any peer-reviewed publications describing results of the recently concluded competition. However, we were able to replicate the methodologies used by Baba using information from his discussion post along with his GitHub repository\footnote{https://github.com/FirasBaba/rsna-ResNet10}. 

The winner experimented with several models before using a surprisingly simple approach with the 3-Dimensional ResNet-10 as the winning solution. It is important to consider how he dealt with the variable slice thickness problem. His solution was to find the slice that contained the largest cross-sectional brain image and treat this as the central slice. This was called the ``Central Image Trick". Then, $n$ images centered around that image were selected as input to the model. The rationale behind this method was that the tumor would be located in the brain area surrounded by the central imaged covered by the selected images. Baba used 64 slices centered around the middle image as the input for his classification CNN model. No further preprocessing was done which indicates that the variable orientations in which the images were provided remained the unchanged when passed into the model. This means that the model would have to learn to predict methylation status from MRI scans provided in various orientations. 

\begin{figure}[H]
     \centering
     \begin{subfigure}[b]{1.8in}
         \centering
         \includegraphics[width=\textwidth]{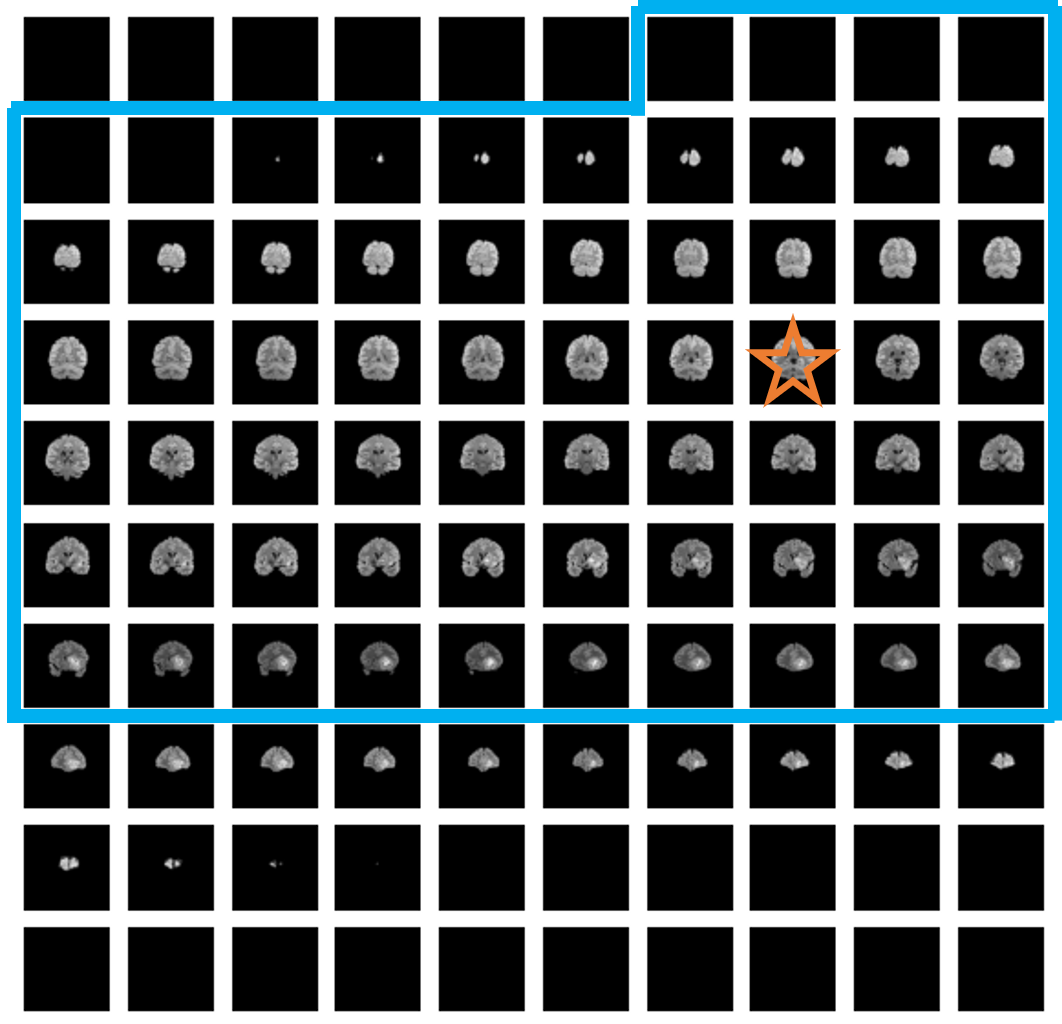}
         \caption{}
         \label{fig:t1}
     \end{subfigure}
     \begin{subfigure}[b]{1.8in}
         \centering
         \includegraphics[width=\textwidth]{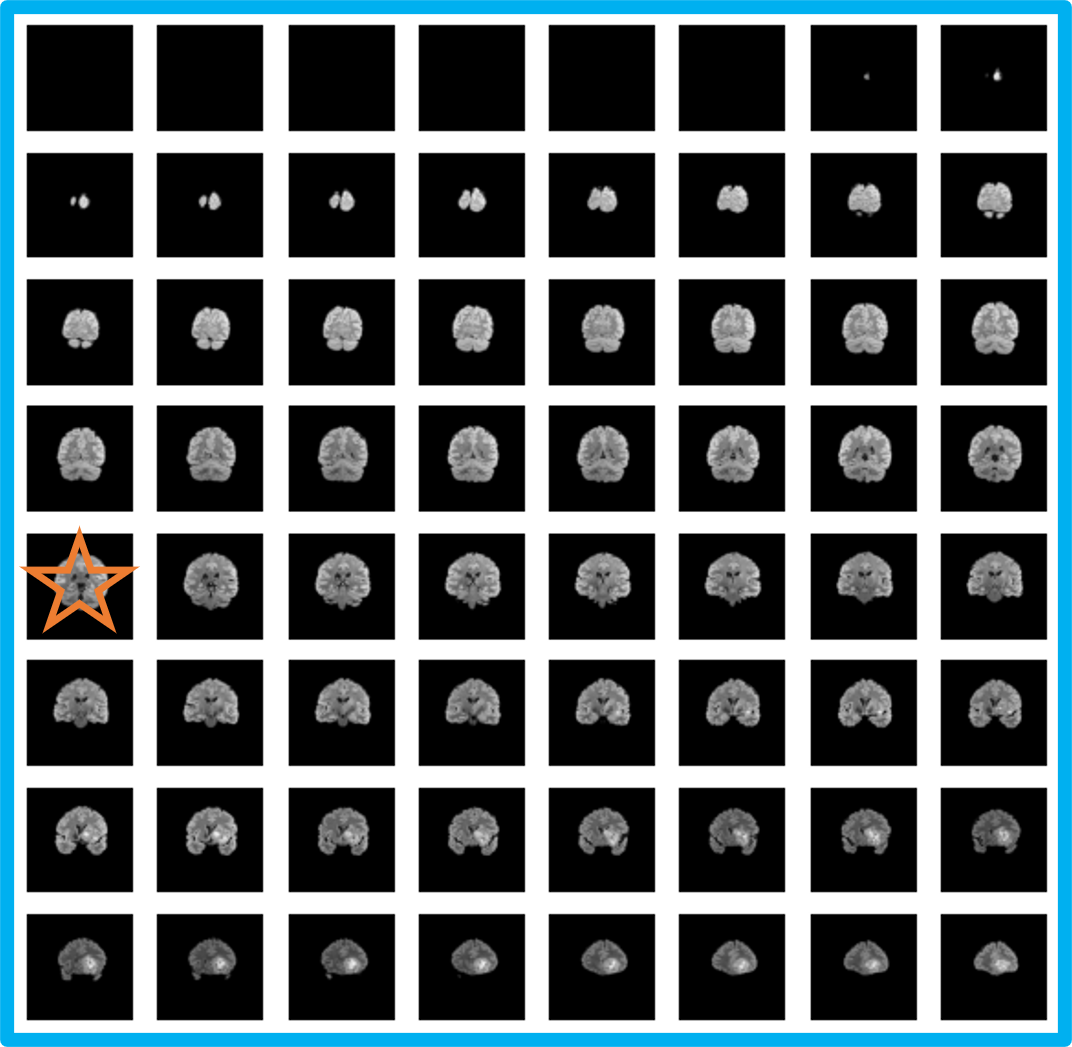}
         \caption{}
         \label{fig:t2}
     \end{subfigure}
     \caption{Standard 64 image selection approach using the Central Image Trick. (a) The central image, marked by the orange star, is identified from the original set of images. (b) 64 images used in classification are selected around the central image.}
     \label{fig:base_select}
\end{figure}

The overall process of selecting the 64 images is a simple two-step process depicted in Figure~\ref{fig:base_select}.

The ResNet-10 was designed using PyTorch and trained for 15 epochs to select the best model used for testing. The results for the baseline ResNet-10 are displayed in the table below.  

\begin{table}[h!]
\centering
\begin{tabular}{||c | c | c | c | c | c||} 
 \hline
 \textbf{Accuracy} & \textbf{Sensitivity} & \textbf{Specificity} & \textbf{PPV} & \textbf{NPV} & \textbf{AUROC} \\
 \hline\hline
 0.60991453 & 0.701628664 &	0.508633094 & 0.615179617 & 0.61794551 & 0.605130879
\\ 
 \hline
\end{tabular}
\caption{Average Cross Validation Results for Baseline ResNet-10 Model.}
\end{table}

These results are consistent with those self-reported by Firas Baba in his blog post.  He obtained average AUROC scores of 0.62 over 5 folds of cross validation. Although his accuracy was at the same level as the enhanced radiomic feature extractor,  he observed a 10\% increase in average specificity.

\subsection{Enhanced ResNet-10}

\tab Although  the baseline ResNet-10 CNN model improves the MGMT classification accuracy for messy MRI data, there are inherent issues with the approach. It is a prevalent clinical practice to examine biopsy tissue sample of the tumor to determine methylation status. Thus, the central issue is that the approach does not rely on the clinical fact that the methylation molecular characteristics are manifested only in the tumor region. The region of brain outside of the tumor should not contain any of these characteristics.  By selecting generic brain regions without any emphasis on tumor regions dilutes the information pertinent for MGMT detection. In many cases, the 64 slices selected as classification input, pivoting on the largest brain image may have little or no tumor region. Additionally, selecting 64 slices ignoring the slice thickness of the scan, results in a different width of the brain being represented for each input. This could misguide the learning process  for the model since it is provided potentially with a different width of brain every training iteration. Finally, the input could be in any of the three spatial orientation forcing  the model to consider an extra variable. 

In the improved enhanced ResNet-10 model, we address the central issue of the baseline model by using only the segmented tumor region as classification input. All non-tumor regions are ignored for this purpose. In order to keep a fixed width of the brain as the input to classification, we normalize the scans to synthetically fill in the intermediary missing slices by harnessing previously described IS-Gen algorithm. Depending on the number of slices provided by the BraTS 2021 dataset for each MRI, we apply the IS-Gen repeatedly to represent the entire brain as a 128x128x128 volume to maintain identical slice thickness for each scan. Thus we create a complete set of MRI scans with uniform slice thickness. In this approach we always maintain a fixed width of the brain when 64 images are selected for classification unlike the baseline model.

We then fully normalize the scans by applying rotational transformation to non-coronal images to coronal plane. Once the scans are normalized, the U-Net based segmentation model as described in Section~\ref{rad_baseline} is used to extract the tumor regions. It is worth mentioning that the segmentation model is trained using BraTS 2020 images in coronal plane. It is expected to perform better since all classification input images are also normalized to the same plane.

\begin{figure}[H]
    \centering
    \includegraphics[width=4in]{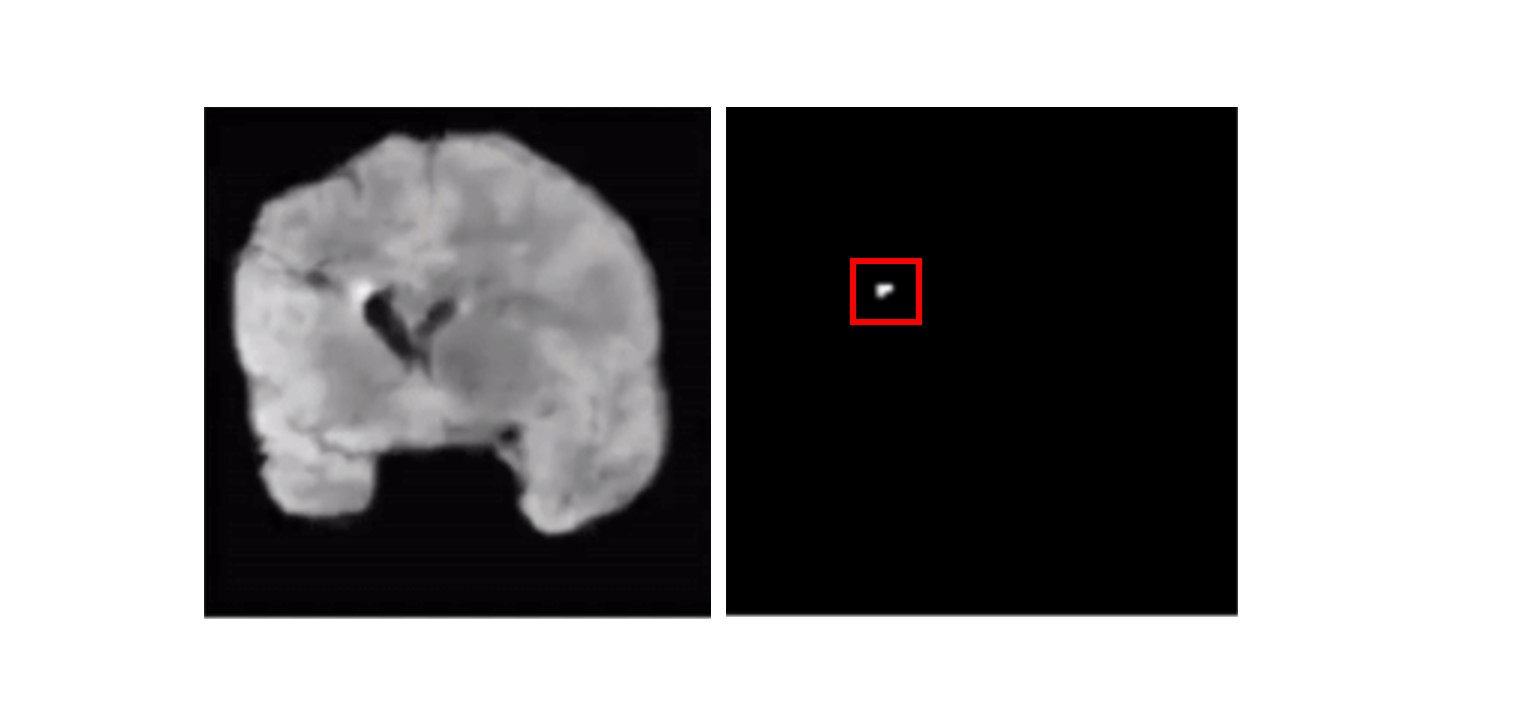}
    \caption{An example false segmentation prediction.}
    \label{fig:bad_seg}
\end{figure}

With segmentation mask for each of the 128 slices in the tumor region of the brain, we need to consider pros and cons of using only a subset of these masks to be used for classification. One can pass in every slice where there is a portion predicted as a tumor region. However, an imperfect segmentation model can cause cases where several slices have small predictions of false tumor regions. Figure~\ref{fig:bad_seg} depicts an example of the segmentation model's imprecise prediction where a non-tumor region has been wrongly predicted as tumor. Using such a slice for classification will deteriorate model's performance.

Alternatively, we can sort each slice based on the size of the predicted mask and choose a subset of the largest tumor regions for classification. However, this could cause a non-continuous region of the brain to be passed in as input. The non-continuous region would result in unrealistic and inaccurate features to be extracted internally by the 3D convolutions.

We make the assumption that tumor region is a single and continuous region of the brain. In order to maintain contiguity, we select the biggest tumor image and then use the 64 images around this center image. We assume that 64 images will adequately capture the entire tumor region without including too much non-tumor  region or excluding too much tumor region for classification. The multistage intelligent tumor selection process is demonstrated in the following Figure~\ref{plot:reset_image_selection}.

\begin{figure}[H]
    \centering
    \captionsetup{width=6in}
    \includegraphics[width=6in]{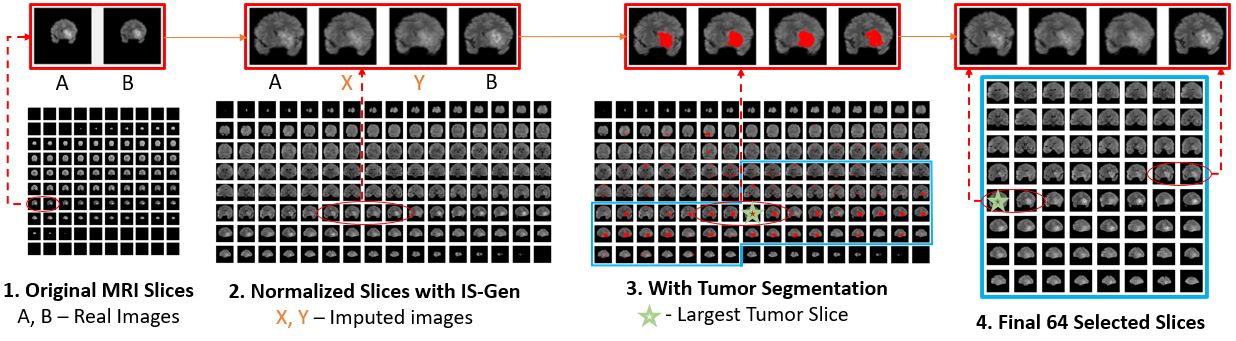}
    \caption{Intelligent image selection procedure used for Enhanced ResNet-10 input. The green star represents the center of the 64 selected images.}
    \label{plot:reset_image_selection}
\end{figure}

Finding the optimal selection of these images is a future area of our research. For example, one could use a sliding window based algorithm to find a continuous subset of images that will cover the tumor region with 95\% confidence.

We used the enhanced ResNet-10 for a direct comparison to the baseline model results. We preprocessed the MRI scans just once before training and stored as \textit{.npy} files to prevent repeated computation. We then trained the model for 40 epochs while saving the model with the highest accuracy after the iteration. We used a NVIDIA K80 GPU to train the model for computational efficiency. As shown in in Table~\ref{table:EnhancedResNet}, we observed more than 6\% increase in overall accuracy relative to the baseline ResNet-10 model.

\begin{table}[H]
\centering
\begin{tabular}{||c | c | c | c | c | c||} 
 \hline
 \textbf{Accuracy} & \textbf{Sensitivity} & \textbf{Specificity} & \textbf{PPV} & \textbf{NPV} & \textbf{AUROC} \\
 \hline\hline
 0.664957265 & 0.783236383 & 0.533116883 & 0.654363102 & 0.746775701 & 0.663826516 \\ 
 \hline
\end{tabular}
\caption{Average Cross Validation Results for Enhanced ResNet-10 Model.}
\label{table:EnhancedResNet}
\end{table}

\section{Results and Discussion}

\begin{figure}[H]
    \centering
    \captionsetup{width=6in}
    \includegraphics[width=6in]{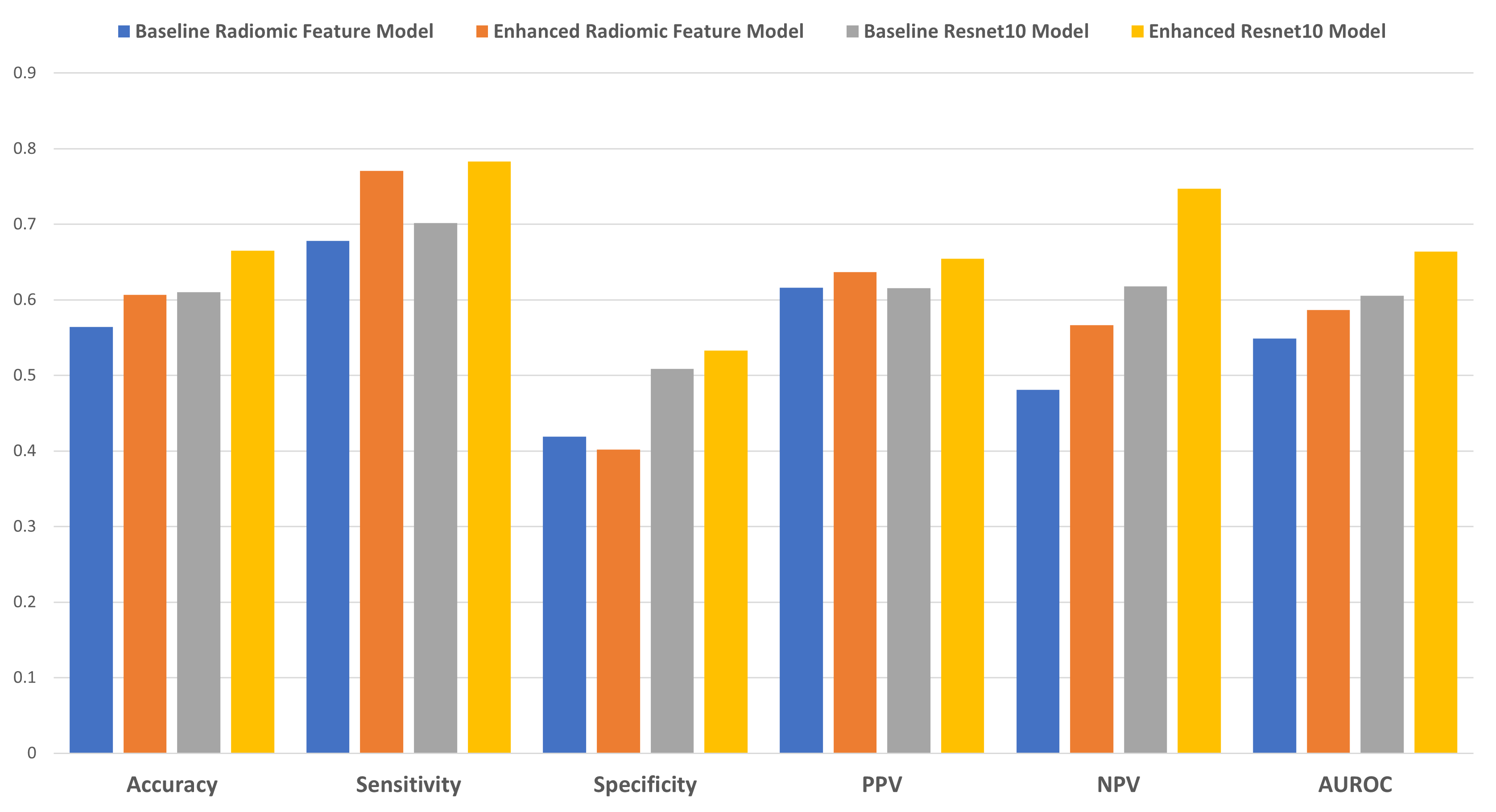}
    \caption{Comparative plot of accuracy, sensitivity, specificity, PPV, NPV, and AUROC scores. }
    \label{plot:performance}
\end{figure}

Finally, we compile model performance results in tabular and graphical formats in Table~\ref{table:performance} and Figure~\ref{plot:performance}, respectively. We notice that our proposed enhanced ResNet-10 model outperforms the other three models we studied in all five performance dimensions. Using Microsoft Excel's \textit{Data Analysis} package, we conducted a 2-way ANOVA with 5 repetitions to compare the enhanced and baseline ResNet-10 models, the top two performing models. The ANOVA result is provided in Table~\ref{table:ANOVA} and confirms that the enhanced ResNet-10 outperforms the Kaggle's best model by a statistically significant margin ($p < 0.05$).


\begin{table}[H]
\centering
\begin{tabular}{||c | c | c | c | c | c | c||} 
 \hline
 & \textbf{Accuracy} & \textbf{Sensitivity} & \textbf{Specificity} & \textbf{PPV} & \textbf{NPV} & \textbf{AUROC} \\
 \hline\hline
 Model 1 & 0.564 & 0.678 & 0.419 & 0.616 & 0.678 & 0.549 \\ 
 \hline
 Model 2 & 0.607 & 0.771& 0.402 & 0.637	& 0.567 & 0.586 \\ 
 \hline
 Model 3 & 0.610 & 0.702 &	0.509 & 0.615 & 0.618 & 0.605 \\ 
 \hline
 Model 4 & 0.665 & 0.783 & 0.533 & 0.654 & 0.747 & 0.664 \\ 
 \hline
\end{tabular}
\caption{Model specific average cross validation results.}
\label{table:performance}
\end{table}

\begin{table}[H]
\centering
\begin{tabular}{||c | c | c | c||} 
 \hline
 \textbf{Source of Variation} & \textit{F} & \textit{P-value} & \textit{F crit} \\
 \hline\hline
 Model Selection & 6.261825405 & 0.015796413 & 4.042652129\\ 
 \hline
\end{tabular}
\caption{2-way ANOVA test results ($\alpha=0.05$).}
\label{table:ANOVA}
\end{table}

 By introducing the improvements in the enhanced model, we ensured that we overcome the shortcomings of the baseline ResNet-10 model. These include normalizing the slice thickness of the brain, maintaining a constant MRI orientation in coronal plane, and intelligently selecting slices containing the tumor region. By reducing the number of variables and enforcing focus on the tumor region only, the enhanced ResNet-10 was able to learn more effectively than the other models. 

For the BraTS 2021 dataset, radiomic feature extraction did not show strong correlations with MGMT promoter methylation. However, we were able to determine that tumor location and shape characteristics were important biomarkers for methylation prediction. Experimenting with the two radiomic feature based models corroborated that the IS-Gen enhances the information content and increases the quality of extracted radiomic features. This was conclusively established with the statistical test described in Section~\ref{enh_radiomic}.

It is worthwhile to note that many different approaches have been reported in literature for classifying MGMT promoter methylation which did not utilize the BraTS 2021 dataset. The reported results for these studies are summarized in Table 10.

\begin{table}[H]
\centering
\begin{tabular}{||p{1in}  p{1.2in}  p{2in}  p{2in} ||} 
 \hline
 \textbf{Author} & \textbf{Performance} & \textbf{Methodology} & \textbf{Data Description} \\
 \hline
 \cite{ahn2014prediction} & Sensitivity - 0.56 
 Specificity - 0.85 
 AUROC - 0.756  & Mann-Whitney and Chi-Square analysis on extracted image biomarkers. & Dynamic contrast-enhanced MRI and Diffusion Weighted Imaging (DWI) from 43 patients. \\ 
 \hline
\cite{han2018mri} & Accuracy - 0.61
Precision - 0.67
Recall - 0.67 & Convolutional Recurrent Neural Network & 159 unique patient MRI scans acquired in 3 modalities from \textit{The Cancer Genome Atlas Glioblastoma Multiforme} (TCGA-GBM) project. All scans were in the Axial plane. Scans contained varying slice thickness. \\ 
 \hline
 \cite{sasaki2019radiomics} & Accuracy - 0.67
 Sensitivity - 0.66 
 Specificity - 0.66 
 PPV - 0.67       
 
 NPV - 0.67 & Radiomic Feature Extraction + LASSO & 181 MRI scans in 3 different modalities collected from 10 institutions. Slice thicknesses for MRI scans were 1 mm.  \\ 
 \hline
\cite{han2018structural} & Accuracy - 0.805
Sensitivity - 0.857
Specificity - 0.852 
AUROC  - 0.877 & Logistic Regression on extracted image biomarkers. & Study used 77 patient data including 3 modalities of MRI scans in the axial orientation with 1.5 mm thickness. Also used DWI scans. \\ 
 \hline
\cite{le2020xgboost} & Accuracy - 0.887
Sensitivity - 0.88
Specificity - 0.887 & Radiomic Feature Extraction + XGBoost Classifier & Used 262 MRI scans of 4 modalities acquired from the TCGA-GBM project. Do not account for varying slice thicknesses.\\ 
 \hline
\end{tabular}

\caption{Summary of existing models that experimented with clean MRI dataset. }
\end{table}

Each author above uses different set of images to determine methylation status. Two of them use DWI which is another form of MRI scans. A few of them make use of the Apparent Diffusion Coefficient (ADC), a value clinically calculated from DWI scans. The ADC is strongly correlated with the MGMT methylation status. Some reported high performance with multimodal MRI scans similar to those provided in the BraTS 2021 dataset. We assume that higher accuracy is a result of using clean nature of the data used for training and classification. For example, \cite{le2020xgboost} used MRI scans acquired from the TCGA-GBM project, a repository of MRI scans labelled with several genetic labels. The authors do not take into account varying slice thicknesses and all images are provided in a constant orientation. 

We attempted to replicate the feature extraction process based on the top 9 radiomic features proposed in this paper. These included first order and GLRLM radiomic values. We did not obtain high performance when applied to our messy dataset. This indicates that the dataset selected could potentially contribute to the level at which MGMT promoter methylation could be learned. The Radiological Society of North America collected images from varying sources without any standard extraction process defined. This can explain the difference in results.


However, for a more consistent comparison to other research, we compared our results to others studies utilizing the BraTS 2021 dataset. Similar to our research~\cite{saeed2021possible} conducted their study after the Kaggle contest was completed in 2021. Not surprisingly, they analyzed the Kaggle winner's solution and made optimizations to the winning solution. 


\begin{table}[H]
\centering
\begin{tabular}{||c | c | c | c | c | c | c||} 
 \hline
 & \textbf{Accuracy} & \textbf{Sensitivity} & \textbf{Specificity} & \textbf{PPV} & \textbf{NPV} & \textbf{AUROC} \\
 \hline\hline
 \cite{saeed2021possible} & - & - & - & - & - & 0.630 \\ 
 \hline
 \cite{palsson2021prediction} & - & - & - & -	& - & 0.598 \\ 
 \hline
 Kaggle Winner & 0.610 & 0.702 &	0.509 & 0.615 & 0.618 & 0.605 \\ 
 \hline
 Proposed Model & \textbf{0.665} & \textbf{0.783} & \textbf{0.533} & \textbf{0.654} & \textbf{0.747} & \textbf{0.664} \\ 
 \hline
\end{tabular}
\caption{Comparison with existing models using the BraTS 2021 dataset. Due to the recent date of the competition, Saeed et al. and Palsson et al. have not been peer reviewed.}
\end{table}

Compared to the performance of models developed by prior research, our proposed model exhibits a significantly higher AUROC score. In future we plan to apply our methodology to the datasets described in Table 10 starting with TCGA-GBM. Due to the fundamental changes in our approach, We expect that our proposed model will work equally well or even better than the other models on clean dataset.

\section{Conclusion}

\tab This study presents a new preprocessing method that improves classification of MGMT promoter methylation directly from MRI scans. We develop a method that can be used to normalize MRI scans into one spatial plane with equal slice thickness by artificially synthesizing missing slices. The proposed method can be beneficial for other tasks aside from MGMT promoter methylation detection. For example, diagnosis of GBM or Alzheimer's Disease can be performed directly from MRI scans~\citep{islam2018early}. With the normalization procedure made possible by the IS-Gen, these methodologies can also be applied to MRI scans produced by a variety of different clinical conditions. The IS-Gen can be adapted for wide range of deep learning models based on MRI scan analysis.

Furthermore, the IS-Gen algorithm is not just limited to MRI scans. It can be applied to any three dimensional dataset to iteratively impute intermediary data layers  to form a more continuous volumetric dataset. For example, with some update to IS-Gen algorithm, it could possibly be used to impute missing frames in a video resulting in higher frames per second. This increases clarity and smoothness of a video. 

Although the IS-Gen proved to contribute additional information, the slices were not always accurate. These inaccuracies caused granular images when reoriented into the coronal plane. We intend to optimize the IS-Gen further by adding more complex CNN feature extractors to increase details of the predicted image. Additionally, we developed a method called On-Off training for the IS-Gen; however, to increase IS-Gen output quality even further, we can design a generator and discriminator to learn at similar rates to remove the need for such training.

In the future, it may be possible to achieve increased performance by applying only the IS-Gen, but removing the reorienting step used in the preprocessing pipelines. This step was done because our segmentation model was trained on MRI scans in the coronal plane. However, with a better segmentation model, we could  use only the IS-Gen to normalize slice thickness and pass in the MRI in it original orientation state. Consequently, we would not lose any segmentation accuracy and images would be crisp. Although this forces the ResNet-10 to distinguish between the distinct orientations, it will likely perform better due to higher image clarity. Finally, we will investigate if improving the performance of the segmentation model further improves prediction accuracy. Specifically, we plan to experiment with the best performing segmentation model from BraTS challenge.

\begin{figure}[H]
    \centering
    \captionsetup{width=6in}
    \includegraphics[width=6in]{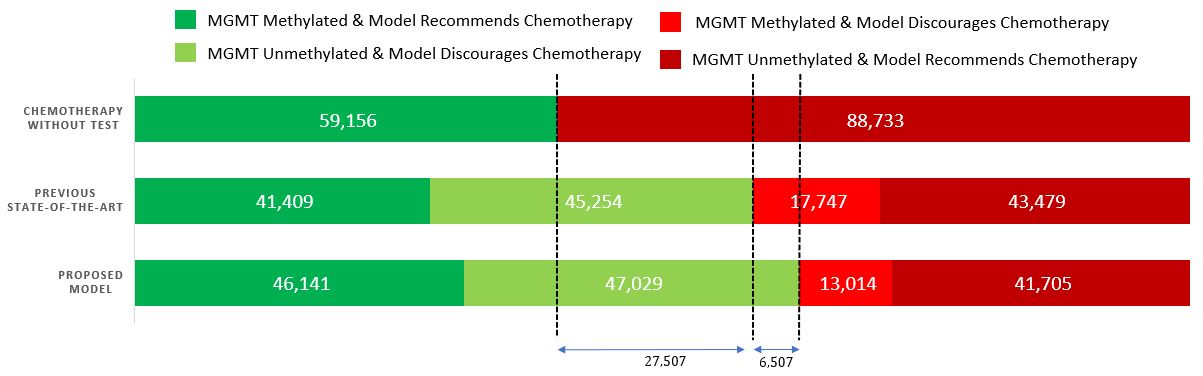}
    \caption{Impact of proposed model when extrapolated to global GBM population.}
    \label{plot:impact}
\end{figure}

Overall, our proposed improvements to the winning Kaggle solution exhibit a 6\% increase in accuracy relative to the prior state-of-the-art. Using statistics provided by the GLOBOCAN 2020 report and sensitivity/specificity values for the baseline and enhanced ResNet-10 models determined by cross-validation, we were able to measure the impact of our proposed model to the global GBM population of approximately 147,889. Specifically, the enhanced ResNet-10 is able to  \textit{correctly recommends} chemotherapy treatment option to 46,141 patients compared to only 41,409 patients by the baseline model. On the contrary, our enhanced model  \textit{correctly discourages} chemotherapy for 47,029 patients compared to only 45,254 patients by the baseline model. In other words, if both models were used to predict cancer treatment in 2020, our proposed model would be able to recommend the correct treatment option to approximately 6,507 more GBM patients than the previous best. If the new model is clinically adopted, it could bring remarkable change in the way GBM patients are treated increasing the lifespan and quality of life for thousands of patients.




\bibliography{bibliography}

\begin{thebibliography}{}

\bibitem[Ahn et~al., 2014]{ahn2014prediction}
Ahn, S.~S., Shin, N.-Y., Chang, J.~H., Kim, S.~H., Kim, E.~H., Kim, D.~W., and
  Lee, S.-K. (2014).
\newblock Prediction of methylguanine methyltransferase promoter methylation in
  glioblastoma using dynamic contrast-enhanced magnetic resonance and diffusion
  tensor imaging.
\newblock {\em Journal of neurosurgery}, 121(2):367--373.

\bibitem[Baid et~al., 2021]{baid2021rsna}
Baid, U., Ghodasara, S., Mohan, S., Bilello, M., Calabrese, E., Colak, E.,
  Farahani, K., Kalpathy-Cramer, J., Kitamura, F.~C., Pati, S., et~al. (2021).
\newblock The rsna-asnr-miccai brats 2021 benchmark on brain tumor segmentation
  and radiogenomic classification.
\newblock {\em arXiv preprint arXiv:2107.02314}.

\bibitem[Bleeker et~al., 2012]{bleeker2012recent}
Bleeker, F.~E., Molenaar, R.~J., and Leenstra, S. (2012).
\newblock Recent advances in the molecular understanding of glioblastoma.
\newblock {\em Journal of neuro-oncology}, 108(1):11--27.

\bibitem[Goodfellow et~al., 2014]{goodfellow2014generative}
Goodfellow, I., Pouget-Abadie, J., Mirza, M., Xu, B., Warde-Farley, D., Ozair,
  S., Courville, A., and Bengio, Y. (2014).
\newblock Generative adversarial nets.
\newblock {\em Advances in neural information processing systems}, 27.

\bibitem[Han and Kamdar, 2018]{han2018mri}
Han, L. and Kamdar, M.~R. (2018).
\newblock Mri to mgmt: predicting methylation status in glioblastoma patients
  using convolutional recurrent neural networks.
\newblock In {\em PACIFIC SYMPOSIUM ON BIOCOMPUTING 2018: Proceedings of the
  Pacific Symposium}, pages 331--342. World Scientific.

\bibitem[Han et~al., 2018]{han2018structural}
Han, Y., Yan, L.-F., Wang, X.-B., Sun, Y.-Z., Zhang, X., Liu, Z.-C., Nan,
  H.-Y., Hu, Y.-C., Yang, Y., Zhang, J., et~al. (2018).
\newblock Structural and advanced imaging in predicting mgmt promoter
  methylation of primary glioblastoma: a region of interest based analysis.
\newblock {\em BMC cancer}, 18(1):1--10.

\bibitem[Islam and Zhang, 2018]{islam2018early}
Islam, J. and Zhang, Y. (2018).
\newblock Early diagnosis of alzheimer's disease: A neuroimaging study with
  deep learning architectures.
\newblock In {\em Proceedings of the IEEE conference on computer vision and
  pattern recognition workshops}, pages 1881--1883.

\bibitem[Kumar et~al., 2012]{kumar2012radiomics}
Kumar, V., Gu, Y., Basu, S., Berglund, A., Eschrich, S.~A., Schabath, M.~B.,
  Forster, K., Aerts, H.~J., Dekker, A., Fenstermacher, D., et~al. (2012).
\newblock Radiomics: the process and the challenges.
\newblock {\em Magnetic resonance imaging}, 30(9):1234--1248.

\bibitem[Le et~al., 2020]{le2020xgboost}
Le, N. Q.~K., Do, D.~T., Chiu, F.-Y., Yapp, E. K.~Y., Yeh, H.-Y., and Chen,
  C.-Y. (2020).
\newblock Xgboost improves classification of mgmt promoter methylation status
  in idh1 wildtype glioblastoma.
\newblock {\em Journal of Personalized Medicine}, 10(3):128.

\bibitem[MacDonald, 2009]{macdonald2009chemotherapy}
MacDonald, V. (2009).
\newblock Chemotherapy: managing side effects and safe handling.
\newblock {\em The Canadian Veterinary Journal}, 50(6):665.

\bibitem[Ostrom et~al., 2014]{ostrom2014cbtrus}
Ostrom, Q.~T., Gittleman, H., Liao, P., Rouse, C., Chen, Y., Dowling, J.,
  Wolinsky, Y., Kruchko, C., and Barnholtz-Sloan, J. (2014).
\newblock Cbtrus statistical report: primary brain and central nervous system
  tumors diagnosed in the united states in 2007--2011.
\newblock {\em Neuro-oncology}, 16(suppl\_4):iv1--iv63.

\bibitem[P{\'a}lsson et~al., 2021]{palsson2021prediction}
P{\'a}lsson, S., Cerri, S., and Van~Leemput, K. (2021).
\newblock Prediction of mgmt methylation status of glioblastoma using radiomics
  and latent space shape features.
\newblock {\em arXiv preprint arXiv:2109.12339}.

\bibitem[Saeed et~al., 2021]{saeed2021possible}
Saeed, N., Hardan, S.~E., Abutalip, K., and Yaqub, M. (2021).
\newblock Is it possible to predict mgmt promoter methylation from brain tumor
  mri scans using deep learning models?

\bibitem[Sasaki et~al., 2019]{sasaki2019radiomics}
Sasaki, T., Kinoshita, M., Fujita, K., Fukai, J., Hayashi, N., Uematsu, Y.,
  Okita, Y., Nonaka, M., Moriuchi, S., Uda, T., et~al. (2019).
\newblock Radiomics and mgmt promoter methylation for prognostication of newly
  diagnosed glioblastoma.
\newblock {\em Scientific reports}, 9(1):1--9.

\bibitem[Savio et~al., 2010]{savio2010effect}
Savio, S.~J., Harrison, L.~C., Luukkaala, T., Heinonen, T., Dastidar, P.,
  Soimakallio, S., and Eskola, H.~J. (2010).
\newblock Effect of slice thickness on brain magnetic resonance image texture
  analysis.
\newblock {\em Biomedical engineering online}, 9(1):1--14.

\bibitem[Sung et~al., 2021]{sung2021global}
Sung, H., Ferlay, J., Siegel, R.~L., Laversanne, M., Soerjomataram, I., Jemal,
  A., and Bray, F. (2021).
\newblock Global cancer statistics 2020: Globocan estimates of incidence and
  mortality worldwide for 36 cancers in 185 countries.
\newblock {\em CA: a cancer journal for clinicians}, 71(3):209--249.

\bibitem[Taal et~al., 2015]{taal2015chemotherapy}
Taal, W., Bromberg, J.~E., and van~den Bent, M.~J. (2015).
\newblock Chemotherapy in glioma.
\newblock {\em CNS oncology}, 4(3):179--192.

\bibitem[Thrower et~al., 2021]{thrower2021effect}
Thrower, S.~L., Al~Feghali, K.~A., Luo, D., Paddick, I., Hou, P., Briere, T.,
  Li, J., McAleer, M.~F., McGovern, S.~L., Woodhouse, K.~D., et~al. (2021).
\newblock The effect of slice thickness on contours of brain metastases for
  stereotactic radiosurgery.
\newblock {\em Advances in Radiation Oncology}, 6(4):100708.

\bibitem[Van~Griethuysen et~al., 2017]{van2017computational}
Van~Griethuysen, J.~J., Fedorov, A., Parmar, C., Hosny, A., Aucoin, N.,
  Narayan, V., Beets-Tan, R.~G., Fillion-Robin, J.-C., Pieper, S., and Aerts,
  H.~J. (2017).
\newblock Computational radiomics system to decode the radiographic phenotype.
\newblock {\em Cancer research}, 77(21):e104--e107.

\end{thebibliography}
\bibliographystyle{apalike}




\end{document}